# Shortest Path Analysis in Social Graphs

Waqas Nawaz[*], Kifayat Ullah Khan, Young-Koo Lee

Dept. of Computer Engineering, Kyung Hee University
Seocheon-dong, Giheung-gu, Yongin-si, Gyeonggi-do, 446-701, Korea
`{wicky786, kualizai, yklee}@khu.ac.kr`

**Abstract.** The shortest path problem is among the most fundamental combinatorial optimization problems to answer reachability queries. It is hard to determine which vertices or edges are visited during shortest path traversals. In this paper, we provide an empirical analysis on how traversal algorithms behave on social networks. First, we compute the shortest paths between set of vertices. Each shortest path is considered as one transaction. Second, we utilize the pattern mining approach to identify the frequency of occurrence of the vertices. We evaluate the results in terms of network properties, i.e. degree distribution, clustering coefficient.

## 1 Introduction

The shortest path problems are among the most fundamental combinatorial optimization problems with many applications, and still remain an active area of research [1]. A single pair shortest path (SPSP) computation using Dijkstra's algorithm requires few seconds on a very large social network even on efficient processing systems [2] [3]. Many traversal algorithms imply pre-computed information to speed up the runtime query process. There is no study available in literature to analyze the occurrence of vertices or edges during the shortest path traversal. In this paper, we provide an empirical analysis on which type of vertices are traversed during shortest path computation in social networks. First, we compute the shortest paths between set of vertices. Each shortest path is considered as one transaction. Second, we utilize the pattern mining approach to identify the frequency of occurrences of the vertices in all transactions.

## 2 Shortest Path Analysis

In this section, we highlight the process of finding frequency of vertex occurrences in real life social graphs. The overall architecture of the system is given in Fig. 1. The source graph data is usually available in the form of set of vertices and edges. There are two main components to find the shortest path overlapped regions. Firstly, the graph traversal algorithm is used to find all pairs of shortest paths, i.e. P. Each path $p_i$ consists of sequence of vertices from source to destination. The graph traversal module produces set of shortest paths between all pair of source and destination as intermediate results. All the shortest paths are computed using well-known Dijkstra algorithm. Secondly, the overlapped regions are identified through pattern mining approach. The visualized statistics enable us to derive useful remarks. The subsequent discussion highlights the details of shortest path computation and pattern mining modules respectively.

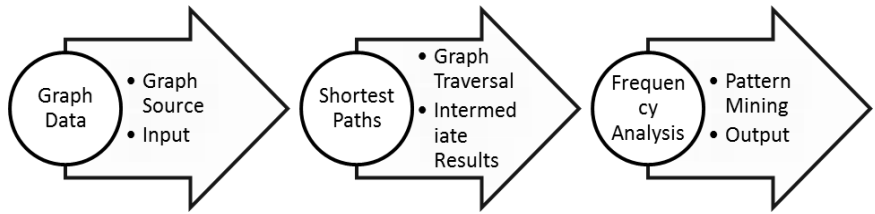

**Fig. 1:** System Architecture

### 2.1 Pair-wise Shortest Path Computation

There are two possibilities to compute the shortest paths between pair of vertices in the social graph, namely exhaustive and non-exhaustive approach. The brute force or straight forward approach is to compute all possible pair of shortest paths from a vertex to another. If N is the number of vertices in the graph then N2 shortest paths are computed. Consequently, N time's single source shortest path (SSSP) algorithm is executed. A single pair shortest path computation in relational environment costs (5M, 15), (10M, 27), (15M, 33), and (20M, 41) in the unit of ($|N|$, sec) [4]. For a huge graph, we randomly choose K vertices and compute K times SSSP for efficient analysis.

### 2.2 Pair-wise Shortest Path Computation

All the shortest paths are reflected as the transaction database. Pattern mining is one of the data mining approaches to find the existing patterns in the data. In this context patterns refer to vertex associations, behaviors, arrangements, or forms in the shortest path. The frequent occurrence of such patterns can lead us to the desired solution. This process of pattern detection is known as frequent pattern mining. There are various approaches in literature to mine frequent patterns. The examples are Apriori, AprioriTID, FP-Growth, Relim, Eclat, and H-mine. The FP-Growth method is efficient compared to other approaches [5]. Using FP-growth, we identify the frequency of one or more vertices in all transactions.

## 3 Experiments

We consider the Facebook dataset from http://snap.stanford.edu/. Fig. 2 shows the degree distribution of the vertices along with the frequency of occurrences for one, two and three consecutive vertices during shortest path computation. We have visualized the original graph along with shortest path traversals for better understanding.

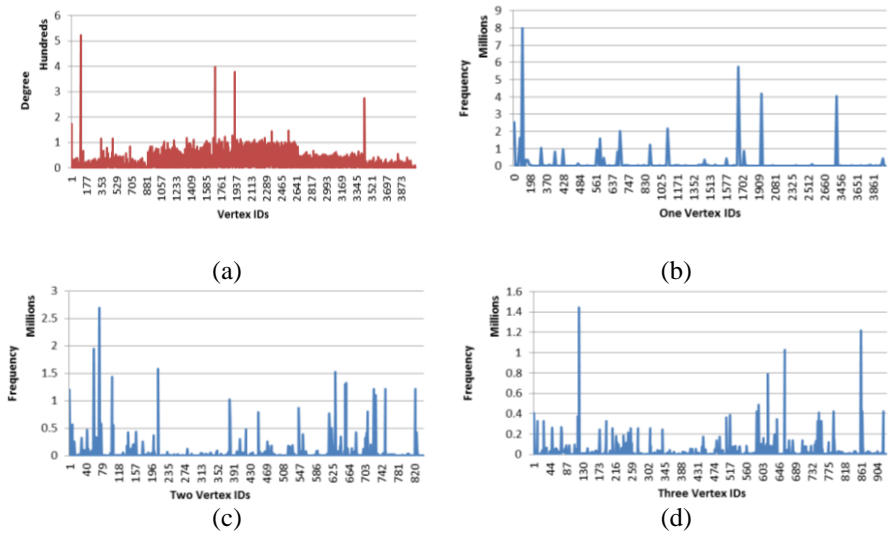

**Fig. 2:** Shortest path analysis (a) degree distribution of vertices, (b) (c) (d) frequency of occurrence for one, two, and three vertices respectively during shortest path computation

## 4 Conclusion and Future Direction

In this paper, we empirically analyze the shortest paths to anticipate the behavior of traversal algorithm on real life network. A set of shortest paths are evaluated using pattern mining approach. We have found that the nodes with very high degree are retained in majority shortest paths. However, nodes with average degree are not considered by the traversal algorithm. Further analysis in terms of network properties, including clustering coefficient, average shortest path, and betweeness centrality, on various types of networks is beyond the scope of this work.

## 5 Acknowledgements

This research was supported by the MSIP (Ministry of Science, ICT&Future Planning), Korea, under the ITRC (Information Technology Research Center) support

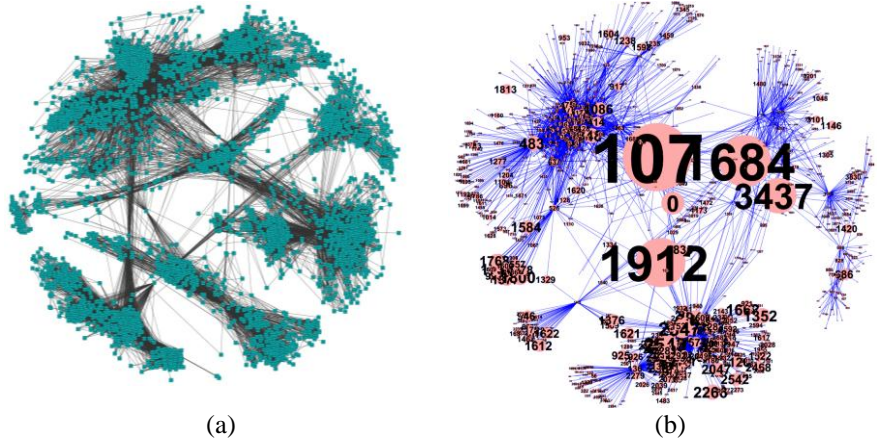

**Fig. 3:** Visualization of the (a) original graph and (b)shortest path traversals